\documentclass[twocolumn,showpacs,preprintnumbers,amsmath,amssymb,prl]{revtex4}
\usepackage{epsfig}
\usepackage{bbm}
\newcommand{\bm}[1]{ \mbox{\boldmath $#1$}  }

\begin{document}

\title{Above threshold $s$-wave resonances illustrated by the 1/2$^+$ states in $^9$Be and $^9$B}

\author{E. Garrido}
\affiliation{Instituto de Estructura de la Materia, CSIC, Serrano 123, E-28006
Madrid, Spain}

\author{D.V. Fedorov and A.S. Jensen}
\affiliation{Department of Physics and Astronomy, University of Aarhus,
DK-8000 Aarhus C, Denmark}
\date{\today}

\begin{abstract}
We solve the persistent problem of the structure of the lowest $1/2^+$
resonance in $^9$Be which is important to bridge the $A=8$ gap in
nucleosynthesis in stars.  We show that the state is a genuine
three-body resonance even though it decays entirely into
neutron-$^8$Be relative $s$-waves.  The necessary barrier is created
by ``dynamical'' evolution of the wave function as the short-distance
$\alpha$-$^5$He structure is changed into the large-distance n-$^8$Be
structure.  This decay mechanism leads to a width about two times
smaller than table values.  The previous interpretations as a virtual
state or a two-body resonance are incorrect. The isobaric analog 1/2$^+$ 
state in $^9$B is found to have energy and width in the vicinity of
2.0 MeV and 1.5 MeV, respectively. We also predict another 1/2$^+$
resonance in $^9$B with similar energy and width.
\end{abstract}


\maketitle

\paragraph{Introduction}

Bound states, resonances and other continuum states are well
understood and described for two particles interacting through a
potential.  By weakening the attraction of the potential bound states
are pushed upwards into the continuum as resonances or virtual states
when a barrier is present or absent, respectively \cite{gar05}.  For
neutral particles virtual states arise for $s$-waves whereas
resonances emerge for higher partial waves.  Decreasing the attraction
further until the resonance energy is above the potential barrier leads to
an increase of the resonance width.  Correspondingly the related
$S$-matrix pole moves in the complex energy plane as a resonance with
non-vanishing imaginary part.

For three particles interacting via two- and three-body potentials the
continuum structures can be much more complicated due to combinations
of the different structures for the three two-body subsystems
\cite{glo96,nie01}.  One intriguing possibility arises when one of the
two-body subsystems has a low-lying $s$-wave resonance produced by a
confining Coulomb barrier, and the third neutral particle has
dominating $s$-wave attractions from the first two particles. Even
when all higher partial waves are vanishingly small, the structure of
the three-body continuum state is a priori not easily determined or
described.

Let us assume that the two-body resonance is very narrow (long-lived)
and the three-body energy is above zero but less than the two-body
resonance energy.  Then the three-body continuum state resembles a
two-body bound state of the third particle and the composite resonance
of the two first particles.  The lifetime would be determined by the
lifetime of the two-body resonance.  When the three-body energy is
pushed upwards above the two-body threshold by decreasing the
attraction, the corresponding structure can be described as a two or
three-body virtual state or resonance \cite{glo96,gar05}.

The purpose of the present letter is first to determine in general
which structure arises, and second specifically to solve the
long-standing controversy of the $^9$Be($1/2^+$) continuum state. This
state is important in the nuclear synthesis of light nuclei in stars
\cite{gor95,arn99,buc01}, and has therefore received lots of attention
both theoretically \cite{gor95,oer96,efr99,bar06} and experimentally
\cite{ang99,sum02,muk05,pap07,bro07,neu09}.  In a measurement of
photo disintegration the cross section is interpreted and parametrized
via $R$-matrix analysis as one neutron and the $^8$Be ground state in
a two-body $s$-wave resonance \cite{sum02}. This seems to be against
the two-body quantum mechanical description as such a state cannot
survive as a resonance.  In another interpretation the same
neutron-$^8$Be system is described as a virtual state \cite{efr99} but
the resulting cross section does not reproduce the measurement
\cite{sum02}.

The $^9$Be($1/2^+$) structure is most often assumed to be one neutron
and the $^8$Be ground state \cite{ang99} but sometimes also the
$\alpha+\alpha+n$ recombination reaction is assumed to proceed by
$\alpha$-capture on the $^5$He ground state \cite{gor95}.  It is
apparently very difficult to avoid assumptions of two-body sequential
structures and processes via subsystems of either $^8$Be or $^5$He.
Interestingly a two-center Born-Oppenheimer model based on symmetries
alone may combine these structures as in \cite{oer96} where the
$1/2^+$ is lowest at large distance whereas a $3/2^-$ state is lowest
at small distance.  We shall allow an entirely general three-body
structure without a priori assumptions of substructures or decay
mechanisms.

\paragraph{Formulation}

Let us consider three composite structures as point-like particles
denoted $n$, $\alpha_1$ and $\alpha_2$. The two mass scaled Jacobi
vector coordinates, $(\bm{x},\bm{y})$, can be substituted by
hyperspherical coordinates $\{\rho,\alpha,\Omega_x,\Omega_y\}$, where
$(\Omega_x,\Omega_y)$ describe the directions of $(\bm{x},\bm{y})$,
$\rho=\sqrt{x^2+y^2}$ and $\alpha=\arctan({x/y})$, see \cite{nie01}.
We solve this three-body problem by use of adiabatic hyperspherical
expansion of the Faddeev equations, i.e.  the angular equations are
solved for each $\rho$, providing a set of angular eigenfunctions 
$\phi_n$ and their corresponding eigenvalues $\lambda_n$.  
The three body wave function is then written as 
$\psi={1\over\rho^{5/2}} \sum_n f_n(\rho) \phi_n(\bm{x},\bm{y})$, where $n$
labels each of the adiabatic terms. The radial functions $f_n(\rho)$ are
obtained after solving a coupled set of radial equations where the 
$\lambda_n$ angular eigenvalues enter as effective adiabatic potentials. 
The coupling between the different adiabatic terms appears through the 
functions $P_{nn'}(\rho)$ and $Q_{nn'}(\rho)$ defined for instance 
in \cite{nie01}.

Each adiabatic potential describes a specific relative structure of
the three particles for a given root mean square radius, $\rho$, with
wave function $\phi$ and eigenvalue $\lambda$. When only one adiabatic
potential is considered, the coupled set of radial equations 
reduces to 
\begin{eqnarray} \label{e40}
\left[ -\frac{d^2}{d\rho^2} + \frac{\lambda(\rho)+ 15/4}{\rho^2} - Q(\rho) -
\frac{2m (E-V_{3b}(\rho) )}{\hbar^2} \right] f(\rho) = 0 \; ,
\end{eqnarray}
where $Q$ is the diagonal coupling term $Q_{nn}$, which 
is in general not zero (contrary to what happens with $P_{nn'}$, whose 
diagonal terms are zero). $V_{3b}(\rho)$ is the three-body potential usually
used in three-body calculations to take into account all those effects that go
beyond the two-body interactions.
The total wave function, $\psi$, and $Q$ are given by
\begin{eqnarray} \label{e50}
\psi={1\over\rho^{5/2}} f(\rho) \phi(\bm{x},\bm{y}) \;\;,
 \;\;  Q(\rho) = \langle \phi | \frac{\partial^2}{\partial \rho^2} | 
 \phi \rangle_{\Omega} \;.
\end{eqnarray}
The expectation value is over angular coordinates, $\Omega$,
excluding only $\rho$.  When only $s$-waves contribute for both $\bm{x}$ and
$\bm{y}$ the angular wave function $\phi$ only depends on $\rho$ and
$\alpha$.  For short-range attractive two-body interactions the
angular eigenvalue $\lambda(\rho)$ would be monotonously increasing
towards a constant asymptotic value.

\paragraph{Narrow two-body resonance.}

\begin{figure}[!h]
\epsfig{file=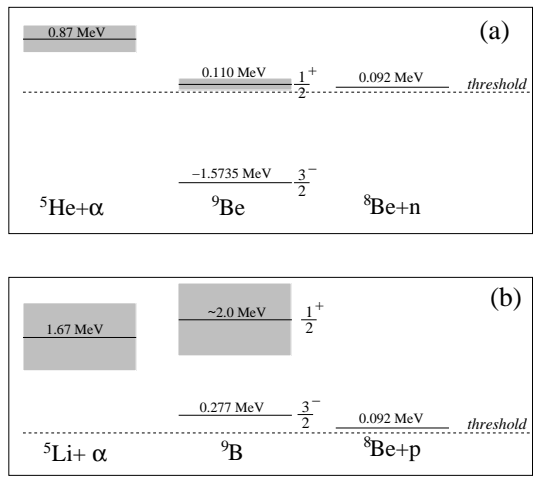, width=7cm, angle=0}
\caption{The two lowest energy levels for $^9$Be (a), and $^9$B (b), and the
resonance energies of the corresponding two-body subsystems \cite{til04}. For $^9$B
the quoted $1/2^+$ state corresponds to the estimation obtained in this work.
The widths of the resonances are represented by the shadowed regions.}
\label{fig0}
\end{figure}

When the $\alpha_1-\alpha_2$ two-body $s$-wave interaction supports a
bound state, the adiabatic potential approaches the bound state energy
at large distance.  This reflects a two-body structure corresponding
to particle $n$ far away from the $\alpha_1-\alpha_2$ bound state.
For less $\alpha_1-\alpha_2$ attraction this state moves into the
continuum. With a confining Coulomb barrier from repelling charges on
the particles, the state would appear as a resonance at low energy
$E_2$. The adiabatic potential would then approach the positive value
equal to $E_2$ corresponding to the large-distance two-body structure
of $n$ far away from the resonance $\alpha_1-\alpha_2$.  We illustrate
in Fig.\ref{fig0} by the specific examples, $^9$Be$(\alpha+\alpha+n)$
and $^9$B$(\alpha+\alpha+p)$.

This description is only correct when the two-body resonance width
$\Gamma_2$ is very small and the coupling to the three-body continuum
states can be ignored.  In general the first adiabatic potential is
crossed by numerous others while $\rho$ increases. However, the
couplings are negligibly small for three-body energies $E$ until
distances $\rho
\simeq 9 \sqrt{E-E_{2}}/\Gamma_2$ where the energies are in MeV and $\rho$ 
in fm.  Thus for small $\Gamma_2$, say eV or keV, the couplings for
moderate energies below $1$~MeV can be neglected far outside the
distance where the short-range $n-(\alpha_1-\alpha_2)$ interaction has
vanished.  Then the system can effectively be described as a two-body
system until the two-body resonance eventually decays.

Let us now consider the three-body system with interactions leading to
a three-body state of positive energy $E_r$ but below $E_2$.
Effectively this is a bound $n-(\alpha_1-\alpha_2)$ two-body state, or
rather a resonance decaying precisely with the width $\Gamma_r
=\Gamma_2$ of the $\alpha_1-\alpha_2$ resonance.  For less
$n-(\alpha_1-\alpha_2)$ $s$-wave attraction this two-body bound state
moves into the continuum above $E_2$. The expectation is that the
proper description is as a virtual state with no width in contrast to
a resonance \cite{efr99}.  However, this is not necessarily true.

\paragraph{The $^9$Be($1/2^+$) example.}

The low-lying states of $^9$Be are expected to be well described as
cluster states consisting of one neutron and two $\alpha$-particles
\cite{cra96}. The $\alpha-\alpha$ interaction, including short range 
attraction and Coulomb repulsion, produces a low-lying $s$-wave
two-body resonance at $0.0918$~MeV with a width around $9$~eV.  Adding
one neutron in $s$-waves leads to angular momentum and parity $1/2^+$
of the resulting three-body system. Such a state is listed in the
tables of energies \cite{til04} at an excitation energy of
$1.684$~MeV, or $0.110$~MeV above threshold, with a width of
$0.217$~MeV (see the upper part of Fig.\ref{fig0}a).  Thus the state
is above the two-body resonance energy by $0.018$~MeV and $s$-waves
are most likely the dominating composition.

The listed width is much larger than the distance to the two-body
resonance threshold and even about two times larger than the
three-body energy itself.  It is a peculiar resonance structure which
apparently extends into the bound state region below the threshold.
These values are consistent with photodissociation cross section
measurements and the entangled $R$-matrix analysis of an a priory
assumed resonance \cite{sum02}.  The fitting parameters are position
and energy dependent width of a two-body neutron-$^8$Be resonance of
$s$-wave character.  Several years prior to this analysis it was
suggested that the state should be understood as a virtual state and
the photodissociation cross section correspondingly analyzed
\cite{efr99}.  However, this does not reproduce the measurements in 
\cite{sum02}.

\begin{figure}[!h]
\epsfig{file=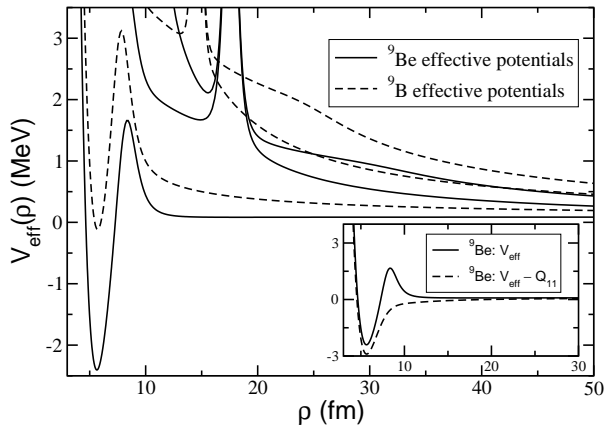, width=7cm, angle=-90}
\caption{Lowest adiabatic potentials for $^9$Be and $^9$B as a function
of the hyperradius. The inset shows the $^9$Be lowest potential 
with and without the rearrangement coupling term $Q$.  }
\label{fig1}
\end{figure}

\paragraph{Three-body results for $^9$Be($1/2^+$).}

The shortcomings of many previous methods are the initial assumption
of two-body character of this structure.  Since the final decay
products are three particles we proceed to treat the system as a
three-body system.  We use the well established nucleon-nucleon and
$\alpha$-nucleon interactions from \cite{alv07,rom08}.  

The adiabatic
potentials produced through the adiabatic hyperspherical expansion
method for these quantum numbers are shown in Fig.\ref{fig1}. 
The higher ones show pronounced peaks at about 15--18 fm. The origin
of the peaks is in the crossing between different angular eigenvalues. 
In particular, the $Q$-functions in the effective potentials, (see
Eqs.(\ref{e40}) and (\ref{e50})), are responsible for the behaviour
shown in the figure in the vicinity of the crossings. These couplings
involve second derivatives of the adiabatic eigenfunctions, and
they therefore reflect restructuring of these functions.
The lowest potential has a dominating attractive pocket at small distance.
After hyperradii $\rho$ larger than $12-14$~fm the potential
stabilizes at the resonance energy of $0.091$~keV for the $^8$Be
ground state.  This stable region continues until interrupted by the
crossings with the (infinitely many) higher potentials. The first 
of these crossings occurs at about $\rho = 130$~fm.

An attractive pocket and a constant large distance potential without
any barrier in the transition region is not able to support a
resonance of finite width at energies above the large distance
asymptotic value.  However, inclusion of the coupling term $Q$ (see
Eq.(\ref{e40})),
provides an otherwise totally absent barrier as seen in the inset of
Fig.\ref{fig1}.  This all decisive barrier then arises from a strong
$\rho$ dependence of the intrinsic (angular) wave function $\phi$, as
seen from the definition in Eq.(\ref{e50}).

\begin{figure}[!h]
\epsfig{file=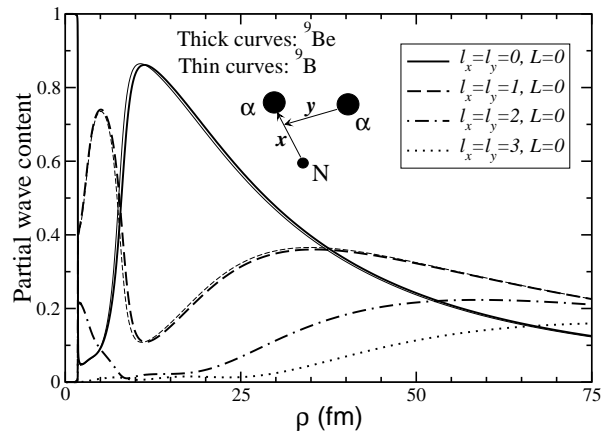,width=7cm, angle=-90}
\caption{The partial wave decomposition of the lowest adiabatic 
angular wave function for $^9$Be (thick) and $^9$B (thin) as function
of hyperradius $\rho$. The partial angular momenta $l_x$ and $l_y$
correspond to the coordinates indicated in the figure. For $l_x=l_y=2$
and $l_x=l_y=3$ the curves for $^9$Be and $^9$B can not be distinguished.  }
\label{fig2}
\end{figure}

The three-body restructuring can be seen from Fig.\ref{fig2} where the
small distance structure of $\rho$ less than $9$~fm is $p$-waves
between neutron and $\alpha$-particles.  This is due to the
$p_{3/2}$-resonance which provides the main part of the attraction.
As $\rho$ increases above $10$~fm this partial wave is rapidly
substituted by $s$-waves which are energetically more advantageous at
larger distance where the attraction vanishes and the centrifugal
barrier dominates.  At much larger distances an increasing number of
partial waves contribute corresponding to the neutron far away from
the two-body system of two $\alpha$-particles in the spatially much
smaller $s$-wave resonance, i.e. $^8$Be in the ground state.

The combined result of the lowest adiabatic potential and the diagonal
coupling is able to support a resonance. We compute the energy and width
numerically as the $S$-matrix pole found by complex scaling \cite{fed03}. 
By adding a structureless short-range potential ($V_{3b}=V_0 \exp(-\rho^2/\rho_0^2)$, 
with $\rho_0=5$~fm), which contributes only in the pocket region, we can  
move the resonance energy by modifying the $V_0$ strength but without
disturbing the resonance structure. This is useful both because the three-body
computation can not place the resonance at precisely the correct
measured energy, and because the width is strongly dependent on the
height and thickness of the barrier at the correct energy.

\begin{figure}[!h]
\epsfig{file=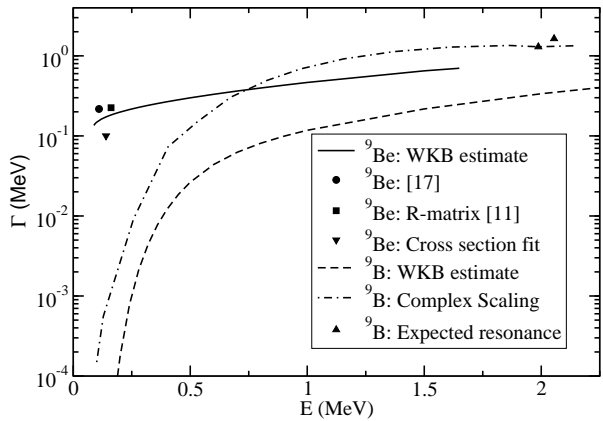, width=7cm, angle=-90}
\caption{Width of the resonances for $^9$Be and $^9$B as function of 
the energy which is varied through the strength $V_0$ of the
three-body potential. The solid and dashed curves are the WKB results with a knocking
rate corresponding to $\Gamma_0=0.6$~MeV ($\Gamma=\Gamma_0 \, e^{-2S}$) for both nuclei 
\cite{gar04}.
The dot-dashed curve results from complex scaling for $^9$B. The square and the circle are
from the R-matrix analysis in \cite{sum02} and the table in \cite{til04}, respectively.
The down triangle is obtained by direct fit of the cross section in \cite{sum02}.
The triangles at about $2$~MeV are the first and second resonances
of $^9$B.  }
\label{fig3}
\end{figure}

In Fig.\ref{fig3} we show the width as a function of the resonance
energy.  The solid line shows the WKB estimate for $^9$Be. For an energy
just above the threshold energy of the $^8$Be ground state ($0.0918$~MeV)
the width is about 0.1 MeV. This agrees with the result obtained by fitting
the measured peak in the photodissociation cross section \cite{sum02} (triangle
down). The width increases slowly up to about $0.7$~MeV
at the top of the barrier. At smaller energies the width is
vanishingly small due to the thick barrier provided by the $^8$Be
structure.

The complex scaling computations for $^9$Be present numerical
difficulties due to the fast increase of the width for energies above
$0.0918$~MeV.  To get an estimate we have instead attached a unit
charge to the neutron, which immediately leads to the results shown in
Fig.\ref{fig3} for the $1/2^+$ analog state in $^9$B.  A continuous
decrease of the charge leads us back to $^9$Be which is approached and
finally reached by extrapolation. Unfortunately the accuracy only
allows the conclusion that the width at an energy of $0.11$~MeV is
larger than $0.06$~MeV and most likely around $0.1$~MeV but $0.2$~MeV
is not numerically excluded. The inaccuracy is due to the very large
width compared to the distance to the threshold.  Then the poles
computed by the complex scaling method are very difficult to
distinguish from the continuum background. This becomes increasingly
worse as the charge of the neutron is decreased from unity to zero
where our present techniques prohibits a clean result.

Comparison to widths obtained in previous works requires precision in
the definitions of a resonance. The ambiguities arise when the resonance
is broad and then necessarily asymmetric. To account for the energy
dependence of the decay probability the $R$-matrix theory employs an
energy dependent width.  This immediately implies that any number
claimed to describe the width must be an average of some kind. The
``observed'' width in $R$-matrix theory is in \cite{til02} most
directly related to the full width at half maximum, or the $S$-matrix
pole, and then adopted as the width in tables of resonance properties
\cite{til04}. The results quoted in \cite{sum02} and \cite{til02} are
shown in Fig.\ref{fig3} by the square and the circle, respectively.

Our computed three-body resonance width is from the WKB tunneling and
crude extrapolations from the imaginary value of the pole of the
$S$-matrix.  Our estimate is very inaccurate but we expect a value of
about $0.1$~MeV which is only about half the ``observed'' table value
from $R$-matrix theory.  This rather large discrepancy can be due to
inaccuracies in the complex scaling extrapolation to zero charge of
the neutron, or the WKB approximation of only one adiabatic potential
combined with the uncertainty in the knocking rate estimate.  However,
we believe that the main reason is that our three-body barrier, which
entirely is responsible for the width, arises from three-body
restructuring effects inherently impossible to include in the
$R$-matrix analysis.  An unambiguous settlement of this accuracy issue
would require the use of another method dedicated to precise width
computations near threshold.

Attempts to settle the issue experimentally face the problems inherent
to relatively very broad resonances.  Population in a reaction or
beta-decay provide information about lifetime, which necessarily is an
average, or details about decay probability as function of energy,
which requires a model for analyzing the data.  The direct
measurements in photo dissociation \cite{sum02} reveal an asymmetric
peak with a width of about $0.1$~MeV.  This is consistent with
corresponding computations in the present model with a method to
compute strength functions by discretization of the three-body
continuum \cite{die08}.

\paragraph{Three-body results for $^9$B($1/2^+$).}

The isobaric analog state should exist in $^9$B but so far it has never
been found. The influence of the additional Coulomb interaction in the
present fragile case could be substantial and perhaps even
destructive.  The corresponding low energy three-body and two-body levels
are shown in Fig.\ref{fig0}b. To discuss this problem we also show in Fig.\ref{fig1}
the corresponding lowest adiabatic potential for $^9$B.  The
additional Coulomb repulsion due to the proton in $^9$B has a relatively
small effect. At small distances the pocket is as pronounced but
almost entirely above zero energy. The barrier is higher and thicker
at small energy where the adiabatic potential itself already exhibits
a small barrier.  The same constant energy is approached at larger
energies again corresponding to $^8$Be in addition to the Coulomb tail
from the proton. The partial wave decomposition in Fig.\ref{fig2} resembles the $^9$Be
results with a tendency towards an increase of higher partial waves
due to the additional Coulomb potential. 

In Fig.\ref{fig3} we show the WKB estimate (dashed line) and the resonance width obtained 
after a complex scaling calculation (dot-dashed line). 
In both cases the width is small even above the $^8$Be ground state energy due to the 
additional adiabatic barrier and the extended Coulomb tail.  
However, the precise value of the energy depends on the three-body potential used in Eq.(\ref{e40}).  
A minimum value for the attractive strength of the three-body force can be obtained 
by placing the $^9$Be resonance at the $^8$Be resonance energy threshold. Using this 
three-body force we can then estimate a lower limit for the 1/2$^+$ resonance in $^9$B, 
which is found to be slightly below 2 MeV with a width of 1.3 MeV (left triangle up in the figure).

This resonance is accompanied by a second state at 2.05 MeV with a
width of 1.6 MeV (right triangle up in the figure).  This second
resonance is very stable, basically independent of the structureless
three-body force.  Therefore, further decrease of the three-body
attraction in $V_{3b}$ would eventually make this second resonance at
2.05 MeV the first 1/2$^+$ excited state.  The two resonances are
dominated by $s$ and $p$-waves between proton and $\alpha$-particle,
respectively.  Crudely speaking these two structures correspond to
$^8$Be$+$proton and $^5$Li$+\alpha$.  Since the large-distance
$p$-wave properties essentially are unaffected by the three-body
potential the second resonance remains close to the same energy.  Thus
the unobserved resonance in $^9$B may in fact be either two or a
combination of two resonances.  In any case we expect genuine
three-body structures with an energy around 2 MeV and a width in the
vicinity of 1.5 MeV.

\paragraph{Discussion and Conclusions.}

The $1/2^+$ continuum structures of $^9$Be and $^9$B are computed as
genuine three-body resonances.  The energies are above the threshold
for forming the ground state $^8$Be-resonance and the partial wave
decompositions are dominated by $s$-waves in the Jacobi coordinates
connecting the two $\alpha$-particles and their center-of-mass and the
nucleon.  Nevertheless the Faddeev component corresponding to the
other Jacobi coordinate presents a very different partial wave
decomposition where the $\alpha$-nucleon $p_{3/2}$ attraction maintain
$p$-waves at hyperradii smaller than $9$~fm rapidly changing into
$s$-waves at $10$~fm.  This ``dynamic evolution'' from $\alpha-^5$He
at small distance to neutron-$^8$Be at intermediate and large distance
reconciles the two limits for the resonance structure.  The
restructuring of the wave function results in a potential barrier
similar to an above barrier reflection at a discontinuity.

The structures are then genuine three-body resonances mixing $s$ and
$p$-waves at small distances while at large distances turning into a
two-body system entirely of $s$-waves for a nucleon and $^8$Be in the
unbound ground state.  The deceiving appearance as a two-body virtual
state is incorrect.  The interpretation as a two-body nucleon-$^8$Be
resonance is also incorrect since the small-distance structure is of
genuine three-body character and this is the very reason for the
existence of a barrier allowing the appearance of resonance features.
Furthermore the parameters from $R$-matrix analysis of the
experimental data is misleading because of the incorrect but crucial
assumption of the existence of a two-body nucleon-$^8$Be resonance.

The astrophysical $n\alpha\alpha$ recombination rate is probably
unaffected provided the corresponding cross section is obtained by
precisely the same parametrization for the measured inverse process of
photodisintegration.  The problem only seems to arise when a different
procedure is applied in these mutually inverse processes. However, all
resonance decays do not necessarily proceed through two-body channels.

The implication is that the decay mechanism is entirely through the
$^8$Be ground state as assumed in most previous publications.
However, the present tabulated value of the resonance width emerges
through an averaging procedure of data analysis and parametrization of
a decaying two-body structure.  The energy dependent width arising
from the $R$-matrix interpretation as a two-body structure is
suspicious since the three-body effects responsible for the barrier
and the width are not included in the analysis.  The true resonance
lifetime should be related to the tunneling probability through the
barrier arising from restructuring the three-body wave function.  The
width is more likely directly found from the peak in the measured
cross section.  We conclude that the controversy over the $1/2^+$
continuum structures of $^9$Be and $^9$B is resolved in a full
three-body model.

\paragraph{Acknowledgments.}

We are grateful for many clarifying discussions with H.O.U. Fynbo and
K. Riisager. This work was partly supported by DGI of MEC (Spain),
contract No. FIS2008-01301.

\end{document}